\documentclass[%
reprint,
superscriptaddress,
bibnotes,
amsmath,amssymb,
aps,
prb,
]{revtex4-2}

\usepackage[main=english,ngerman]{babel}
\usepackage[T1]{fontenc}
\usepackage[utf8]{inputenc}
\usepackage{lmodern}
\usepackage{microtype}
\usepackage{graphicx}
\usepackage[hidelinks]{hyperref}
\usepackage{amsfonts}
\usepackage{siunitx}
\usepackage{physics}
\usepackage{mathtools}
\usepackage{cleveref}
\usepackage{xfrac}
\usepackage{placeins}

\begin{document}

\title{Tunable WS\texorpdfstring{\textsubscript{2}}{2} Micro-Dome Open Cavity Single Photon Source}

\author{Jens-Christian Drawer}%
\affiliation{\begin{otherlanguage}{ngerman}Carl von Ossietzky Universität Oldenburg, Fakultät V, Institut für Physik, 26129 Oldenburg\end{otherlanguage}, Germany}%

\author{Salvatore Cianci}%
\affiliation{Physics Department, Sapienza University of Rome, 00185 Rome, Italy}%

\author{Vita Solovyeva}%
\affiliation{\begin{otherlanguage}{ngerman}Carl von Ossietzky Universität Oldenburg, Fakultät V, Institut für Physik, 26129 Oldenburg\end{otherlanguage}, Germany}%

\author{Alexander Steinhoff}%
\affiliation{\begin{otherlanguage}{ngerman}Carl von Ossietzky Universität Oldenburg, Fakultät V, Institut für Physik, 26129 Oldenburg\end{otherlanguage}, Germany}%

\author{Christopher Gies}%
\affiliation{\begin{otherlanguage}{ngerman}Carl von Ossietzky Universität Oldenburg, Fakultät V, Institut für Physik, 26129 Oldenburg\end{otherlanguage}, Germany}%

\author{Falk Eilenberger}%
\affiliation{Friedrich Schiller University Jena, Institute of Applied Physics, 07745 Jena, Germany}%

\author{Kenji Watanabe}%
\affiliation{Research Center for Electronic and Optical Materials, National Institute for Materials Science, 1-1 Namiki, Tsukuba 305-0044, Japan}%

\author{Takashi Taniguchi}%
\affiliation{Research Center for Materials Nanoarchitectonics, National Institute for Materials Science, 1-1 Namiki, Tsukuba 305-0044, Japan}%

\author{Ivan Solovev}%
\affiliation{\begin{otherlanguage}{ngerman}Carl von Ossietzky Universität Oldenburg, Fakultät V, Institut für Physik, 26129 Oldenburg\end{otherlanguage}, Germany}%

\author{Giorgio Pettinari}%
\affiliation{Institute for Photonics and Nanotechnologies, National Research Council, 00133 Rome, Italy}%

\author{Federico Tuzi}%
\affiliation{Physics Department, Sapienza University of Rome, 00185 Rome, Italy}%

\author{Elena Blundo}%
\affiliation{Physics Department, Sapienza University of Rome, 00185 Rome, Italy}%

\author{Marco Felici}%
\affiliation{Physics Department, Sapienza University of Rome, 00185 Rome, Italy}%

\author{Antonio Polimeni}%
\affiliation{Physics Department, Sapienza University of Rome, 00185 Rome, Italy}%

\author{Martin Esmann}%
\affiliation{\begin{otherlanguage}{ngerman}Carl von Ossietzky Universität Oldenburg, Fakultät V, Institut für Physik, 26129 Oldenburg\end{otherlanguage}, Germany}%

\author{Christian Schneider}%
\thanks{Corresponding author:\\ \href{mailto:christian.schneider@uni-oldenburg.de}{christian.schneider@uni-oldenburg.de}}
\affiliation{\begin{otherlanguage}{ngerman}Carl von Ossietzky Universität Oldenburg, Fakultät V, Institut für Physik, 26129 Oldenburg\end{otherlanguage}, Germany}%

\date{\today}%

\begin{abstract}
	Versatile, tunable, and potentially scalable single-photon sources are a key asset in emergent photonic quantum technologies. In this work, a single-photon source based on WS\textsubscript{2} micro-domes, created via hydrogen ion irradiation, is realized and integrated into an open, tunable optical microcavity.	Single-photon emission from the coupled emitter--cavity system is verified via the second-order correlation measurement, revealing a $g^{(2)}(\tau=0)$ value of \num{0.3}. A detailed analysis of the spectrally selective, cavity enhanced emission features shows the impact of a pronounced acoustic phonon emission sideband, which contributes specifically to the non-resonant emitter--cavity coupling in this system. The achieved level of cavity--emitter control highlights the potential of open-cavity systems to tailor the emission properties of atomically thin quantum emitters, advancing their suitability for real-world quantum technology applications.
\end{abstract}

\maketitle

\section{Introduction}
Compact and efficient sources of quantum light are of critical importance in establishing quantum communication networks and are advancing as building blocks in on-chip quantum information processing. In recent years, a large palette of solid-state single-photon emitters has emerged, featuring different degrees of flexibility and versatility. While InAs-based quantum dots still perform best in terms of single-photon emission brightness and purity, atomically thin semiconductors are promising candidates as a versatile, scalable, and low-cost alternative. Thus far, single-photon emission was demonstrated from localized exciton emitters in strained transition-metal dichalcogenide (TMDC) mono- and bilayers \cite{srivastava_optically_2015, koperski_single_2015, he_single_2015, chakraborty_voltage-controlled_2015}, He$^+$ ion beam--irradiated MoS\textsubscript{2} monolayers \cite{klein_engineering_2021, barthelmi_atomistic_2020}, as well as moiré-trapped excitons in van der Waals heterostructures \cite{baek_highly_2020}. In a recent work, the feasibility of creating micro-domes via hydrogen ion beam irradiation of thin TMDC thin flakes was demonstrated, which represents a novel degree of freedom in engineering both the global and the local strain in van der Waals heterostructures \cite{tedeschi_controlled_2019, blundo_giant_2025}. It was furthermore demonstrated that such micro-domes can host single-photon emitters, rendering them a unique approach towards deterministic sources of quantum light \cite{cianci_tailoring_2022, cianci_spatially_2023}.

In this work, we realize a cavity-tunable single-photon source based on a WS\textsubscript{2} micro-dome realized via hydrogen ion irradiation. We demonstrate quantum light emission from the deterministically coupled emitter--cavity system, and, via combining spectrally selective emitter--cavity coupling with an adapted model accounting for phonon degrees of freedom, we explore the fingerprints arising from the strong phonon sidebands of the 2D material coupled to the optical resonator mode \cite{steinhoff_impact_2025, vannucci_single-photon_2024}.

\section{Methods and results}
\begin{figure*}
	\includegraphics{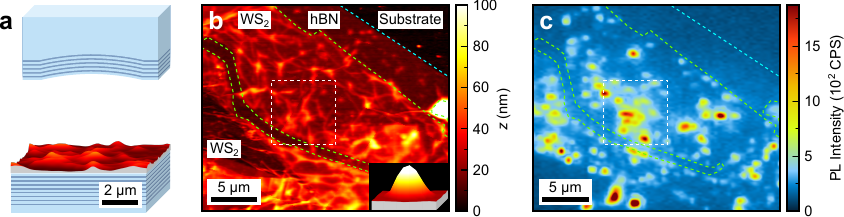}%
	\caption{\label{fig:fig1}%
		(a)~To-scale schematic of the $\sim$\SI{5}{\micro m} long open cavity containing the hydrogenated and hBN-capped WS\textsubscript{2} layer. The topography of the active layer is scaled by a factor of \num{10} in the $z$ direction.
		(b)~AFM measurement of the hBN-capped, hydrogenated WS\textsubscript{2} layer. The edges of the WS\textsubscript{2} layer are indicated by a dashed green line, one of the edges of the hBN layer on top of it is shown by the dashed blue line, and the part shown in (a) is indicated by a dashed white square. An inset at the bottom right shows a dome before it was capped with hBN. It is scaled by a factor of \num{1.5} compared to (a).
		(c)~PL map matching the region shown in (b), displaying the spectrally integrated intensity in the \num{622}--\SI{650}{nm} range under \SI{532}{nm} continuous wave laser excitation.}%
\end{figure*}

The device consists of a hydrogenated WS\textsubscript{2} flake, which is integrated into a spectrally tunable, open Fabry--Pérot cavity (see \cref{fig:fig1}a). The lower mirror is a distributed Bragg reflector (DBR) made of \num{10.5} pairs of TiO\textsubscript{2}/SiO\textsubscript{2}, terminated with a SiO\textsubscript{2} layer, with optical layer thicknesses chosen as $\sfrac{\lambda}{4}$, such that the optical stop band is centered at \SI{620}{nm}. A nominally \SI{5}{nm} thick flake of WS\textsubscript{2}, exfoliated from the bulk crystal by the Scotch-tape method, is transferred to the mirror using the dry-gel stamping technique. This step is followed by the low-energy hydrogen beam irradiation procedure described in \cite{blundo_engineered_2020, tedeschi_controlled_2019}, which results in the formation of one-atomic-monolayer-thick domes that evolve at the surface of the flake. Finally, the sample is capped with an approximately \SI{8}{nm} thick layer of hexagonal boron nitride (hBN), making the domes stable even at cryogenic temperatures \cite{cianci_spatially_2023}.

The top mirror consists of \num{5.5} layers of TiO\textsubscript{2}/SiO\textsubscript{2}, and lateral optical mode confinement is provided by a spherical-cap-shaped indentation (diameter of \SI{6}{\micro m}, depth of \SI{330}{nm}). More details can be found in Section~I of the Supporting Information.

First, we characterized the properties of the active hydrogenated WS\textsubscript{2} before mounting the cavity top mirror. Atomic force microscopy (AFM) measurements of the hBN-capped, hydrogenated WS\textsubscript{2} layer are shown in \cref{fig:fig1}b and the inset of \cref{fig:fig1}b displays a dome before hBN capping. While the uncapped dome features a highly symmetric shape, the topography on the hBN surface is strongly modified by formation of wrinkles in the capping layer. \Cref{fig:fig1}c displays a spatially resolved micro-photoluminescence (PL) map of the studied flake, recorded at a sample temperature of \SI{3.9}{K} using a confocal microscopy configuration without the top mirror. The sample flake is excited non-resonantly via an expanded \SI{532}{nm} continuous wave diode laser. The spectrally integrated intensity in the emission window \num{622}--\SI{650}{nm} is displayed, where we expect the luminescence from localized emitters in WS\textsubscript{2} domes \cite{cianci_spatially_2023}. We find that the PL emitted from the sample is highly anisotropic and localized; spot-like emission arises at positions that we assign to the occurrence of micro-domes.

\begin{figure*}
	\includegraphics{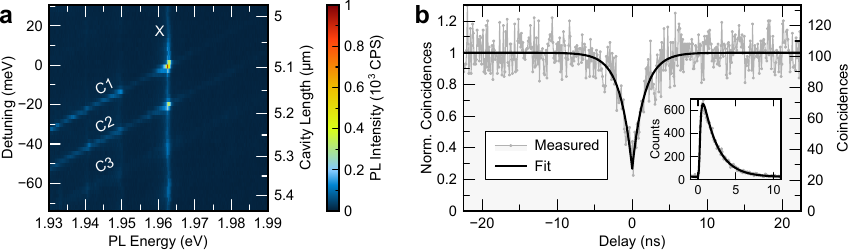}%
	\caption{\label{fig:fig2}%
		(a)~PL spectra for various detunings between cavity resonance energy and emitter energy, denoted as X, at approximately \SI{1.962}{eV}. The diagonal features correspond to the spectrally tunable cavity modes, labeled C1--C3. On resonance (zero detuning), the intensity is enhanced by approximately a factor of \num{17}.
		(b)~Second-order correlation measurement at zero detuning, yielding $g^{(2)}(0)=\num{0.27\pm0.08}$. The inset shows a lifetime measurement of the coupled system.}%
\end{figure*}

Cavity coupling of single emission centers is accomplished by mounting the top distributed Bragg reflector in our open cavity framework. Since we can laterally align the two mirrors with nanometer-scale precision, it is possible to deterministically place a pre-selected emitter underneath the center of the spherical-cap-shaped indentation in the top DBR. \Cref{fig:fig2}a displays a map of PL spectra for \SI{532}{nm} continuous wave laser excitation that were recorded as the vertical distance between top and bottom DBRs was continuously tuned in a range between \num{4.98} and \SI{5.44}{\micro m}. The primary, narrow-band emission signal of the localized emitter remains spectrally stable at an energy of \SI{1.962}{eV} (labeled as X in \cref{fig:fig2}a). For negative cavity--emitter detunings (i.e., $E_\textrm{c}-E_\textrm{x}<0$, where $E_\textrm{c}$ is the energy of the cavity resonance, and $E_\textrm{x}$ is the energy of the localized exciton), non-resonant emitter--cavity coupling \cite{ates_non-resonant_2009} gives rise to a weaker, but distinct, luminescence signal of the optical resonances of our cavity, with a linewidth of $\sim$\SI{2.3}{meV} (labeled as C1 in \cref{fig:fig2}a). As we change the cavity length, the resonant mode can be fully tuned through the discrete, localized emission line, yielding a significant enhancement of the detected PL signal, up to a factor of approximately \num{17}. We also detect higher-order cavity modes with energies \SI{20}{meV} and \SI{42}{meV} above the ground mode (labeled as C2 and C3 in \cref{fig:fig2}a, respectively), corresponding to higher-order Laguerre-Gaussian modes trapped in the plano-concave micro-cavity \cite{bennenhei_polarized_2023, han_infrared_2025}. The higher-order modes display less pronounced coupling behavior, since C2 features a field node in the lateral center of the photonic trap, and both C2 and C3 exhibit increased scattering losses compared to C1 due to their larger mode diameters. It is worth pointing out that, for positive detunings ($E_\textrm{c}-E_\textrm{x}>0$), the emission from the cavity mode is strongly reduced as a result of a strong spectral asymmetry in the acoustic phonon sideband \cite{mitryakhin_engineering_2024}, which we investigate more closely later in this paper.

Single-photon emission from the resonantly coupled emitter--cavity system (zero detuning conditions) is verified by measuring the second-order correlation in a Hanbury-Brown and Twiss (HBT) setup under continuous wave and non-resonant excitation. We use a coarse spectral bandpass to filter the \num{600}--\SI{650}{nm} window containing the primary emission feature, and couple the signal into a single mode fiber via a zoom collimator to allow optimal mode matched coupling. For the HBT measurement, a 50:50 fiber beam splitter is used with each output connected to an avalanche photodiode, both of which are connected to a time correlation device. The resulting correlation histogram, shown in \cref{fig:fig2}b, clearly displays a pronounced dip at zero time delay, which is the hallmark of non-classical light emission. The fit of the model $g^{(2)}(\tau)=1-p\cdot e^{-\abs{\tau}/\tau_1}$, convoluted with the system response function, yields $g^{(2)}(0)=\num{0.27\pm0.08}$, verifying the single-photon nature of the emission from our device. The inset of \cref{fig:fig2}b shows a lifetime measurement of the emitter for pulsed \SI{532}{nm} excitation (\SI{8}{\micro W} focused intensity, $\sim$\SI{5}{ps} pulse duration). We model the setup response with a skewed Gaussian function and deconvolute the data by fitting the ideal model convoluted by the setup response function, yielding a characteristic decay time of $\tau_1=\SI{1.954\pm0.024}{ns}$ (see Section~II~C of the Supporting Information for details).

\begin{figure*}
	\includegraphics{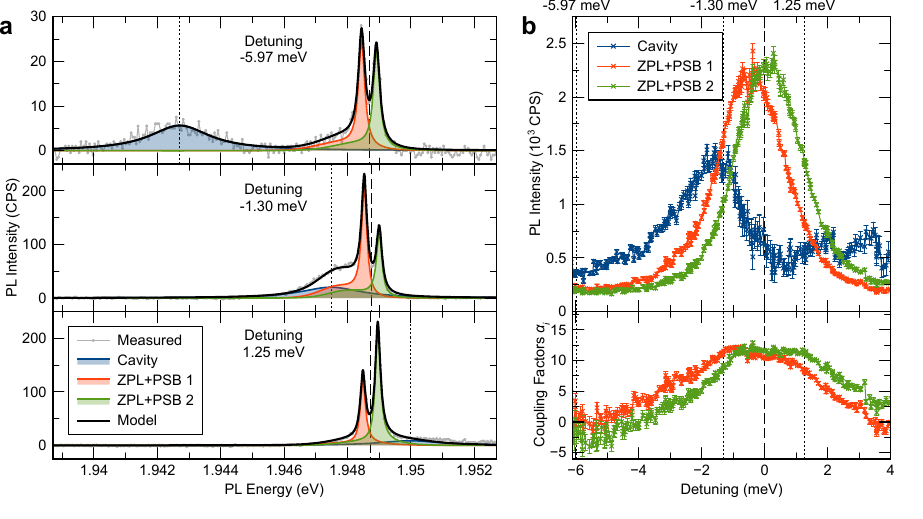}%
	\caption{\label{fig:fig3}%
		(a)~Measured PL spectra for three cavity emitter detunings with the fitted model, as well as its individual constituents, namely, the cavity and the lower (higher) energy emission feature ``ZPL+PSB 1'' (``ZPL+PSB 2'').
		(b)~Integrated intensities attributed to the three modeled components (top) and the cavity emitter coupling factors (bottom) as a function of the cavity emitter detuning. The vertical thin dashed lines correspond to the three spectra shown in (a).}%
\end{figure*}

The highly tunable nature of our open cavity system allows us to gain further insight into the intricate interplay between the resonant coupling between cavity, quantum emitter, and phonon modes. To this end, we selected another emitter from a different dome, featuring a pronounced coupling with the lattice vibrations. In \cref{fig:fig3}a, we display high-resolution emission spectra of a selected emitter for strong (top panel) and modest (middle panel) negative detunings, as well as slightly positive detuning (bottom panel). The emission feature of this selected emitter is split into a doublet of discrete emission lines (zero-phonon lines, ZPLs), resulting from the inherent fine-structure of the apparent charge-neutral, localized exciton. As mentioned, the emitter has a pronounced acoustic phonon sideband (PSB) that emerges from phonon emission events at energies below the ZPL. As in the discussion of \cref{fig:fig2}a, we notice that the emission of the cavity mode is clearly visible for modest negative detuning, while it carries a substantially lower intensity at positive detuning, since phonon absorption events are strongly reduced at the chosen sample temperature of \SI{4}{K}. For a complete map of PL spectra for the vertical distance tuning, refer to Figure~S3 in the Supporting Information. Our second-order correlation measurement verifies single-photon emission for this selected emitter, and the result can be found in Figure~S4 in the Supporting Information.

In order to carry out a quantitative analysis, we have extended the model that was introduced in \cite{abbarchi_phonon_2008} to describe the two individual ZPL-PSB constituents of the spectra by the influence of the cavity as detailed in \cite{mitryakhin_engineering_2024}. In this approach, the spectral shapes of the ZPL $I_\textrm{ZPL}(E)$ and of the cavity mode $I_\textrm{c}(E)$ are described by Lorentzians, and the acoustic-phonon sideband (PSB) contribution is given by \cite{abbarchi_phonon_2008}
	\begin{equation}
	\begin{split}
		I_\textrm{PSB}(E)={}&a\exp\!\left[-\left(\frac{E-E_\textrm{x}}{\gamma_\textrm{ph}}\right)^{\!2}\right]\times\abs{E-E_\textrm{x}}\\
		&\times\left(\Theta(E_\textrm{x}-E)+\frac{1}{\exp\!\left(\frac{\abs{E-E_\textrm{x}}}{k_\textrm{B}T}\right)-1}\right),
	\end{split}
\end{equation}
where the temperature is $T=\SI{4}{K}$.

For spectra with a significantly red-detuned cavity, any interaction beyond a simple additive superposition of the constituents is neglected, and a fit of type
\begin{equation}
	I(E) = \sum_{i=1,2}\left(I_{\textrm{ZPL},i}(E)+I_{\textrm{PSB},i}(E)\right) + I_\textrm{c}(E) + c
\end{equation}
can be performed for each individual spectrum to extract the intensities, energies, and linewidth of the cavity mode, as well as the quantum emitter features. The index $i=1$ ($i=2$) corresponds to the lower (higher) energy emission feature of ZPL and PSB. Finally, the entire set of spectra for different cavity detunings is described with an individually fitted set of the remaining parameters, that is, the width of the ZPL Lorentzians $I_{\textrm{ZPL},i}$, their average center energy $\frac{E_{\textrm{x},1}+E_{\textrm{x},2}}{2}$, and the amplitude of the cavity feature $I_\textrm{c}$. To this end, the adapted model
\begin{equation}
	\begin{split}
		I(E)=&\sum_{i=1,2}\left(I_{\textrm{ZPL},i}(E)+I_{\textrm{PSB},i}(E)\right)\cdot\left[1+\alpha_iI'_\textrm{c}(E)\right]\\
		&+ I_\textrm{c}(E) + c\label{eq:fullMainText}
	\end{split}
\end{equation}
is employed, where $I'_\textrm{c}$ is the normalized cavity mode intensity $I'_\textrm{c}\equiv\frac{I_\textrm{c}}{I_\textrm{c}(E=E_\textrm{c})}$. The expression given by \cref{eq:fullMainText} accounts for the leaky emission of the two lines for far detuned cavity energies and describes how the cavity preferentially enhances the emission of the line that lies closer to its resonance. In contrast to \cite{mitryakhin_engineering_2024}, two additional coupling factors $\alpha_i$ are introduced as fit parameters, which are shown in \cref{fig:fig3}b (bottom). They capture the deviation from the cavity acting as a purely multiplicative filter on the emission lines, allowing for better reproduction of the complete cavity length scan and improving the extraction accuracy of the individual components, whose integrated intensities are then given by
\begin{align}
	I_\textrm{c,Int}=&\int_{-\infty}^\infty I_\textrm{c}(E)\dd{E},\\
	I_{\textrm{ZPL+PSB},i,\textrm{Int}}=&\int_{-\infty}^\infty \left(I_{\textrm{ZPL},i}(E)+I_{\textrm{PSB},i}(E)\right)\nonumber\\
	&\phantom{\int_{-\infty}^\infty}\times\left[1+\alpha_iI'_\textrm{c}(E)\right]\dd{E},\quad i=1,2,
\end{align}
and which are shown in \cref{fig:fig3}b (top). From matching the measured spectra to our model, we obtain an average linewidth for the ZPL Lorentzians of $\textrm{FWHM}_{\textrm{ZPL},i}=\SI{0.2054\pm0.0004}{meV}$. For an extensive step-by-step description of the fitting procedure, refer to Section~III of the Supporting Information.

The central result of our data analysis reveals that the phonon-sideband emission funnels into the optical cavity mode, yielding the peculiar, strongly asymmetric intensity distribution of the cavity that peaks at slightly negative detunings. Interestingly, while the intensity of the ZPL reaches its maximum at zero-detuning, effectively, the intensity of the cavity mode undergoes a minimum, which is analogous to findings for single InAs quantum dots coupled to micropillar cavities \cite{suffczynski_origin_2009} but which has not yet been observed for van der Waals quantum emitters.

\section{Conclusion}
Our work demonstrates a micro-dome single-photon source in a tunable, open optical cavity. We explore the main spectral fingerprints of the micro-dome quantum emitters and their coupling behavior to the resonator. In particular, we find pronounced phonon-sideband emission, which yields significant non-resonant emitter--cavity coupling under negative detuning conditions, similar to observations in WSe\textsubscript{2} emitters in wrinkles \cite{mitryakhin_engineering_2024}. Furthermore, we observe a strong quenching of the cavity mode intensity when the system is on resonance. Since micro-dome quantum emitters can be generated deterministically in TMDC flakes, our work outlines the possibility for a scalable implementation of spatially and spectrally deterministic TMDC-based single-photon sources. The pronounced exciton-phonon coupling also makes micro-dome single-photon sources interesting candidates for applications in quantum-optomechanics \cite{barzanjeh_optomechanics_2022, kettler_inducing_2021}.

\begin{acknowledgments}
	This project was funded by the \begin{otherlanguage}{ngerman}Deutsche Forschungsgemeinschaft\end{otherlanguage} {} (DFG, German Research Foundation), grant numbers INST 184/222-1 FUGG, INST 184/234-1 FUGG, INST 184/235-1 FUGG, and Gi1121/4-2. This project was funded within the QuantERA II programme that has received funding from the European Union’s Horizon 2020 research and innovation programme under Grant Agreement No.\ 101017733, and with funding organization the Germany Federal ministry of research, technology and aeronautics within the project EQUAISE. The \begin{otherlanguage}{ngerman}Niedersächsisches Ministerium für Wissenschaft und Kultur\end{otherlanguage} {} within the collaborative project DyNano is acknowledged. We gratefully acknowledge funding by the BMFTR within the project Tublan (FKZ 16KISQ089 and FKZ 16KISQ088). AP and MF acknowledge funding from the PNRR MUR project PE0000023-NQSTI (the National Quantum Science and Technology Institute). KW and TT acknowledge support from the JSPS KAKENHI (Grant Numbers 21H05233 and 23H02052), the CREST (JPMJCR24A5), JST and World Premier International Research Center Initiative (WPI), MEXT, Japan. The authors thank Heiko Knopf for his assistance with sample fabrication. The authors thank Ilaria Rago and Francesco Pandolfi for support with the AFM measurements.
\end{acknowledgments}

\bibliography{zotero}

\end{document}


\title{\texorpdfstring{Supporting Information\\Tunable WS\texorpdfstring{\textsubscript{2}}{2} Micro-Dome Open Cavity Single Photon Source}{Supporting Information -- Tunable WS\texorpdfstring{\textsubscript{2}}{2} Micro-Dome Open Cavity Single Photon Source}}

\author{Jens-Christian Drawer}%
\affiliation{\begin{otherlanguage}{ngerman}Carl von Ossietzky Universität Oldenburg, Fakultät V, Institut für Physik, 26129 Oldenburg\end{otherlanguage}, Germany}%

\author{Salvatore Cianci}%
\affiliation{Physics Department, Sapienza University of Rome, 00185 Rome, Italy}%

\author{Vita Solovyeva}%
\affiliation{\begin{otherlanguage}{ngerman}Carl von Ossietzky Universität Oldenburg, Fakultät V, Institut für Physik, 26129 Oldenburg\end{otherlanguage}, Germany}%

\author{Alexander Steinhoff}%
\affiliation{\begin{otherlanguage}{ngerman}Carl von Ossietzky Universität Oldenburg, Fakultät V, Institut für Physik, 26129 Oldenburg\end{otherlanguage}, Germany}%

\author{Christopher Gies}%
\affiliation{\begin{otherlanguage}{ngerman}Carl von Ossietzky Universität Oldenburg, Fakultät V, Institut für Physik, 26129 Oldenburg\end{otherlanguage}, Germany}%

\author{Falk Eilenberger}%
\affiliation{Friedrich Schiller University Jena, Institute of Applied Physics, 07745 Jena, Germany}%

\author{Kenji Watanabe}%
\affiliation{Research Center for Electronic and Optical Materials, National Institute for Materials Science, 1-1 Namiki, Tsukuba 305-0044, Japan}%

\author{Takashi Taniguchi}%
\affiliation{Research Center for Materials Nanoarchitectonics, National Institute for Materials Science, 1-1 Namiki, Tsukuba 305-0044, Japan}%

\author{Ivan Solovev}%
\affiliation{\begin{otherlanguage}{ngerman}Carl von Ossietzky Universität Oldenburg, Fakultät V, Institut für Physik, 26129 Oldenburg\end{otherlanguage}, Germany}%

\author{Giorgio Pettinari}%
\affiliation{Institute for Photonics and Nanotechnologies, National Research Council, 00133 Rome, Italy}%

\author{Federico Tuzi}%
\affiliation{Physics Department, Sapienza University of Rome, 00185 Rome, Italy}%

\author{Elena Blundo}%
\affiliation{Physics Department, Sapienza University of Rome, 00185 Rome, Italy}%

\author{Marco Felici}%
\affiliation{Physics Department, Sapienza University of Rome, 00185 Rome, Italy}%

\author{Antonio Polimeni}%
\affiliation{Physics Department, Sapienza University of Rome, 00185 Rome, Italy}%

\author{Martin Esmann}%
\affiliation{\begin{otherlanguage}{ngerman}Carl von Ossietzky Universität Oldenburg, Fakultät V, Institut für Physik, 26129 Oldenburg\end{otherlanguage}, Germany}%

\author{Christian Schneider}%
\thanks{Corresponding author:\\ \href{mailto:christian.schneider@uni-oldenburg.de}{christian.schneider@uni-oldenburg.de}}
\affiliation{\begin{otherlanguage}{ngerman}Carl von Ossietzky Universität Oldenburg, Fakultät V, Institut für Physik, 26129 Oldenburg\end{otherlanguage}, Germany}%

\date{\today}%

\maketitle

\renewcommand{\thefigure}{S\arabic{figure}}
\renewcommand{\theequation}{S\arabic{equation}}

\section{\label{sec:SIsample}Sample preparation}
\begin{figure*}
	\includegraphics{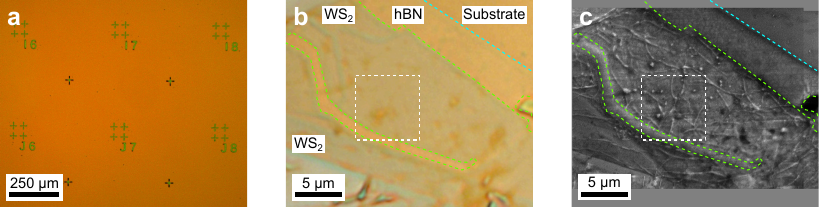}%
	\caption{\label{fig:figS2}%
		(a)~Light microscope image of the gold alignment markers patterned on the lower DBR mirror, forming a reference grid used for flake localization.
		(b)~Light microscope image of the hBN-capped hydrogenated WS\textsubscript{2} layer of the same section and with the same labels as shown in fig.~1b and fig.~1c in the main text.
		(c)~Scanning electron microscope image corresponding to the region in (b). Note that this image is assembled from two separate images. It does not fill the entire field of view.}%
\end{figure*}

The open cavity consists of two opposing mirrors, forming a Fabry--Pérot resonator. The lower mirror is a distributed Bragg reflector (DBR), composed of \num{10.5} pairs of TiO\textsubscript{2}/SiO\textsubscript{2} layers, terminated with a SiO\textsubscript{2} layer. The optical thicknesses of the individual layers are chosen as $\sfrac{\lambda}{4}$, resulting in a stop band centered at \SI{620}{nm}. Gold alignment markers (Cr/Au) are patterned on the mirror via electron-beam lithography, metal evaporation, and lift-off. These markers form a reference grid that enables reliable localization of individual flakes, even when the sample is mounted inside the cryostat and imaged with a simple optical microscope. A light microscope image of the finished marker pattern is shown in \cref{fig:figS2}a. An approximately \SI{5}{nm} thick flake of WS\textsubscript{2}, exfoliated from the bulk crystal by the Scotch-Tape method, is transferred onto the mirror using the dry-gel stamping technique. This step is followed by the low-energy ($\sim$\SI{20}{eV}) hydrogen beam irradiation procedure described in \cite{blundo_engineered_2020, tedeschi_controlled_2019}, which results in the formation of atomically thin, one mono-layer-thick domes appearing uniformly on the surface of the crystal. Indeed, given the thinness of the flake, the sample is irradiated with a low dose of \SI{8e14}{protons/cm^2}, leading to the appearance of modestly sized domes. Their topography is measured using atomic force microscopy (AFM, Park Systems NX10 AFM) in tapping mode, with a representative result shown in the inset of fig.~1b of the main text. Finally, the sample is capped with an approximately \SI{8}{nm} thick layer of hBN, making the domes stable even at cryogenic temperatures of a few Kelvins \cite{cianci_spatially_2023, blundo_giant_2025} by preserving their strained configuration, even when the condensed hydrogen is no longer able to exert the necessary pressure. The adherence of the hBN layer is favored by the low density of domes on the crystal surface, providing large areas of flat WS\textsubscript{2} on which the hBN can effectively anchor itself. The topography of the fully assembled sample is measured again using AFM (WITec alpha300 R) in tapping mode, with the results shown in fig.~1b of the main text. A light microscope image and a scanning electron microscope (SEM, JEOL JSM-IT800) image of the same region, corresponding to fig.~1b and fig.~1c in the main text, can be found in \cref{fig:figS2}b and \cref{fig:figS2}c, respectively.

\begin{figure*}
	\includegraphics{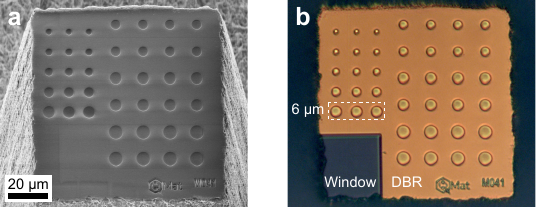}%
	\caption{\label{fig:figS3}%
		(a)~SEM image of the mesa used as the top mirror of the open cavity after FIB milling.
		(b)~Light microscope image of the mesa after DBR coating and fabrication of a transparent window by milling away the coating in a planar region of the mesa. The lens size used in the measurements is indicated by a dashed box.}%
\end{figure*}

The top mirror is made of SiO\textsubscript{2}, into which a $\SI{100}{\micro m}\times\SI{100}{\micro m}$ sized, \SI{100}{\micro m} high protrusion (mesa) is milled by laser writing. A pattern of spherical-cap-shaped indentations of \num{3}, \num{3.5}, \num{4}, \num{5}, \num{6}, and \SI{7}{\micro m} diameter and \SI{330}{nm} depth is milled into it using a focused Ga$^+$ ion beam (FIB, FEI Helios 600i). An SEM image (FEI Helios 600i) of the FIB milled mesa is shown in \cref{fig:figS3}a. Finally, \num{5.5} layers of TiO\textsubscript{2}/SiO\textsubscript{2} are sputtered onto it, again with optical thicknesses chosen as $\sfrac{\lambda}{4}$ to obtain a distributed Bragg reflector, with its stopband centered at $\lambda=\SI{680}{nm}$. To facilitate alignment of the top and bottom mirrors, the mirror coating is partially removed, again with the FIB, in a planar region of the mesa. A light microscope image of the finished mesa is shown in \cref{fig:figS3}b.

\section{Experimental methods}
\subsection{Micro PL measurements}
The pre-characterization of the samples includes micro-PL mapping of the studied flake, with data shown in fig.~1c in the main text. It is performed without the top mirror of the cavity installed inside a cold finger cryostat (Attocube AttoDry 800) at \SI{3.9}{K} temperature and using a confocal microscopy configuration. The sample flake is globally illuminated by a \SI{532}{nm} continuous wave diode laser, which is expanded by focusing into the back focal plane of the $\textrm{NA}=\num{0.82}$ cold objective (Attocube LT-APO/VISIR/\num{0.82}). The PL is imaged onto the entrance slit of a spectrometer (Andor Shamrock SR-750i), equipped with a Peltier-cooled charge coupled device (CCD, Andor iKon-M 934 series) as a detector. This allows to assemble a spectrally resolved map of the sample's emission by moving it stepwise perpendicular to the slit (along the $x$ axis), while collecting 2D spectra with the spatial $y$ coordinate resolved along the slit and the $x$ axis on the CCD chip spectrally resolved.

\subsection{Open cavity spectroscopy measurements}
For the assembled open cavity, a set of $xyz$ piezoelectric nanopositioners is used for each mirror, allowing integration into a liquid helium-free closed-cycle cryostat operating at \SI{3.2}{K} (Attocube AttoDry 1000). A scale drawing of the assembled open cavity with a typically used \SI{5}{\micro m} cavity length can be found in fig.~1a in the main text. The optical setup is again given by a confocal microscopy setup where excitation and collection are performed through the indented top mirror of the device. In this case, an Andor Shamrock SR-500i spectrometer is used. More details can be found in \cite{drawer_monolayer-based_2023, mitryakhin_engineering_2024}, where an identical optical setup was used.

\subsection{\label{sec:SIcorr}Single-photon correlation measurements}
The single-photon line is spectrally bandpass filtered to the \num{600}--\SI{650}{nm} window. For both the second-order correlation measurement in the HBT configuration and the lifetime measurement, the signal is then coupled into a single mode fiber via a zoom collimator (Thorlabs ZC618APC-A), to allow optimal mode matched coupling. For the HBT measurement, a 50:50 fiber beam splitter is used, with each output connected to an avalanche photodiode (APD, Laser Components COUNT T-Series) as single-photon detectors, which are connected to a time correlation device (Swabian Time Tagger 20). For the lifetime measurement, a pulsed \SI{532}{nm} laser source (NKT SuperK EXTREME with NKT SuperK SELECT, with additional \SI{532}{nm} bandpass and \SI{550}{nm} shortpass filters) with a pulse duration of about \SI{5}{ps} is used. The start signal is provided electronically by the fast photodiode integrated in the laser source, while the stop signal is provided by a fiber-coupled APD. In both cases, HBT and lifetime, the bin sizes are set to \SI{100}{ps} width, and the integration runs until \num{100} counts per bin are exceeded. The system response function of both setups is taken into account for the extraction of $\tau_1$ and $g^{(2)}(0)$ by performing measurements for the collection of the pulsed \SI{532}{nm} source in both setups, modeling the result with a skewed Gaussian, and deconvoluting the data by fitting the ideal model convoluted by the modeled system response. The timing jitter of a single APD is specified by the manufacturer to be in the range of \SI{500}{ps}. In the case of the HBT measurement, the data shown in fig.~2b in the main text is the measured histogram, normalized to the value extracted for $g^{(2)}(\tau\rightarrow\infty)$.

\section{\label{sec:SIfitting}Modeling the emitter in a tuned cavity}
In order to carry out a quantitative analysis, we have extended the model that was introduced in \cite{abbarchi_phonon_2008} to describe the two individual ZPL-PSB constituents of the spectra by the influence of the cavity as detailed in \cite{mitryakhin_engineering_2024}. The three constituents are:
\begin{enumerate}
	\item The acoustic-phonon sideband (PSB) contribution is \cite{abbarchi_phonon_2008}
	\begin{equation}
		\begin{split}
			I_\textrm{PSB}(E)={}&a\exp\!\left[-\left(\frac{E-E_\textrm{x}}{\gamma_\textrm{ph}}\right)^{\!2}\right]\times\abs{E-E_\textrm{x}}\\
			&\times\left(\Theta(E_\textrm{x}-E)+\frac{1}{\exp\!\left(\frac{\abs{E-E_\textrm{x}}}{k_\textrm{B}T}\right)-1}\right),
		\end{split}
	\end{equation}
	where the temperature is $T=\SI{4}{K}$. Fits are done with respect to the emitters center energy $E_\textrm{x}$, the Gaussian width $\gamma_\textrm{ph}$, and the intensity of the feature $a$.
	\item The ZPL intensity $I_\textrm{ZPL}(E)$ is described by a Lorentzian, fitted with respect to amplitude and width; its center energy matches $E_\textrm{x}$ of the PSB.
	\item The cavity mode intensity $I_\textrm{c}(E)$ is modeled by a Lorentzian as well and fitted with respect to height and width. Its center energy $E_\textrm{c}$ is preset for each spectrum by extracting it from a smooth curve fitted to the cavities energy shift from a plot like in fig.~2a in the main text.
\end{enumerate}
For spectra with a significantly red-detuned cavity, any interaction beyond a simple additive superposition of the constituents is neglected, and a fit of type
\begin{equation}
	I(E) = \sum_{i=1,2}\left(I_{\textrm{ZPL},i}(E)+I_{\textrm{PSB},i}(E)\right) + I_\textrm{c}(E) + c
\end{equation}
can be performed for each individual spectrum. The index $i=1$ ($i=2$) corresponds to the lower (higher) energy emission feature of ZPL and PSB. Note that for all fits emission energies above \SI{1.951}{eV} are neglected to avoid components unrelated to the two emission lines. The following properties are extracted as inverse-variance weighted averages of the fitted series for the subset of cavity energies in the \num{1.9413}--\SI{1.9426}{eV} range, as here the integrated intensities $\int_{-\infty}^\infty I_{\textrm{ZPL},i}(E)+I_{\textrm{PSB},i}(E)\dd{E}$ are lowest.
\begin{itemize}
	\item The energy spacing $E_{\textrm{x},2}-E_{\textrm{x},1}=\SI{0.4715\pm0.0021}{meV}$ of the two emission line features.
	\item The amplitude of the ZPL Lorentzian $I_{\textrm{ZPL},i}$, which is set to be identical for $i=1,2$ in each fit.
	\item The width $\gamma_\textrm{ph}=\SI{1.20\pm0.04}{meV}$ and amplitude $a$ of the PSB $I_{\textrm{PSB},i}$, which are also set to be identical for $i=1,2$ in each fit.
	\item The width of the cavity feature $I_\textrm{c}$, giving $\textrm{FWHM}_\textrm{c}=\SI{2.30\pm0.07}{meV}$.
	\item A constant background $c$, which is also subtracted from the measured data shown in fig.~3a in the main text.
\end{itemize}
Their errors are obtained from propagating individual fit errors through the inverse-variance weighted average expression. While not fixed to the weighted average here, one obtains an average linewidth of the ZPL Lorentzians $I_{\textrm{ZPL},i}$ of $\textrm{FWHM}_{\textrm{ZPL},i}=\SI{0.227\pm0.004}{meV}$, which were also set to be identical for $i=1,2$ in the fits. Finally, the entire set of spectra for different cavity detunings is described with an individually fitted set of the remaining parameters, that is, the width of the ZPL Lorentzians $I_{\textrm{ZPL},i}$, their average center energy $\frac{E_{\textrm{x},1}+E_{\textrm{x},2}}{2}$, and the amplitude of the cavity feature $I_\textrm{c}$. To this end, the adapted model
\begin{align}
		I(E)=&\sum_{i=1,2}\left(I_{\textrm{ZPL},i}(E)+I_{\textrm{PSB},i}(E)\right)\cdot\left[1+\alpha_iI'_\textrm{c}(E)\right]\nonumber\\
		&+ I_\textrm{c}(E) + c\label{eq:full}
\end{align}
is employed, where $I'_\textrm{c}$ is the normalized cavity mode intensity $I'_\textrm{c}\equiv\frac{I_\textrm{c}}{I_\textrm{c}(E=E_\textrm{c})}$. The expression given by \cref{eq:full} accounts for the leaky emission of the two lines for far detuned cavity energies and describes how the cavity preferentially enhances the emission of the line that lies closer to its resonance. In contrast to \cite{mitryakhin_engineering_2024}, two additional coupling factors $\alpha_i$ are introduced as fit parameters, which are shown in fig.~3b (bottom) in the main text. They capture the deviation from the cavity acting as a purely multiplicative filter on the emission lines, allowing for better reproduction of the complete cavity length scan and improving the extraction accuracy of the individual components, whose integrated intensities are then given by
\begin{align}
	I_\textrm{c,Int}=&\int_{-\infty}^\infty I_\textrm{c}(E)\dd{E},\\
		I_{\textrm{ZPL+PSB},i,\textrm{Int}}=&\int_{-\infty}^\infty \left(I_{\textrm{ZPL},i}(E)+I_{\textrm{PSB},i}(E)\right)\nonumber\\
		&\phantom{\int_{-\infty}^\infty}\times\left[1+\alpha_iI'_\textrm{c}(E)\right]\dd{E},\quad i=1,2,
\end{align}
and which are shown in fig.~3b (top) in the main text. From matching the measured spectra to our model, we obtain an average linewidth for the ZPL Lorentzians of $\textrm{FWHM}_{\textrm{ZPL},i}=\SI{0.2054\pm0.0004}{meV}$.

\FloatBarrier
\begin{widetext}
\section{Additional data}
\begin{figure*}[h]
	\includegraphics{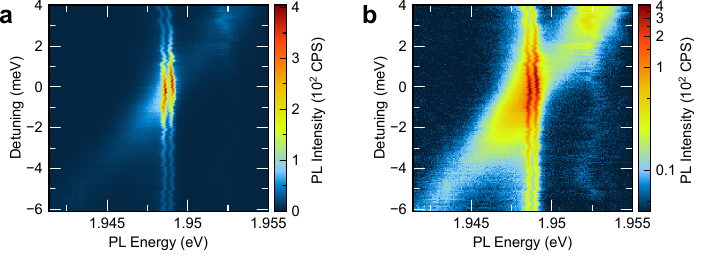}%
	\caption{\label{fig:figS1}%
		PL spectra corresponding to the data shown in fig.~3 in the main text for various detunings between cavity resonance energy and ZPL energy at approximately \SI{1.949}{eV}. (a)~Linearly and (b)~logarithmically scaled intensity false color-scale. The diagonal feature corresponds to the cavity mode.}%
\end{figure*}

\begin{figure*}[h]
	\includegraphics{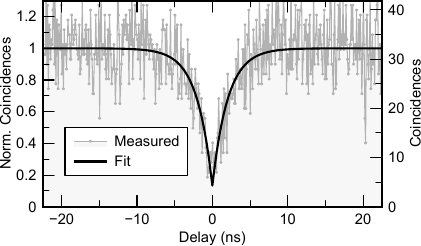}%
	\caption{\label{fig:figS4}%
		Second-order correlation measurement for the emitter used to analyze its spectral contributions in fig.~3 in the main text. The measurement conditions are otherwise identical to those employed in fig.~2b in the main text, but a higher excitation intensity of \SI{30}{\micro W} is used, and the integration is stopped at $\sim$35 counts per bin at large delays. The fit of the model $g^{(2)}(\tau)=1-p\cdot e^{-\abs{\tau}/\tau_1}$, convoluted with the system response function, gives $g^{(2)}(0)=\num{0.13\pm0.12}$ and $\tau_1=\SI{2.0\pm0.4}{ns}$.}%
\end{figure*}
\FloatBarrier
\end{widetext}

\bibliography{zotero}